**Interlayer Pores Play a Limited Role in Diffusion Through Hydrated Na-MMT: Insights from a Multiscale, Experimentally Anchored Model**

Yaoting Zhang[1], Mikaella Brillantes[2], Justine Kuczera[1], Keyvan Ferasat[1], Mia L. San Gabriel[2], Scott Briggs[3], Chang Seok Kim[3], George Opletal[4], Yuankai Yang[5], Jane Howe[2] and Laurent K. Beland[1*]

[1] Department of Mechanical and Materials Engineering, Queen's University, Kingston, Canada.

[2] Department of Materials Science and Engineering, University of Toronto, Toronto, Canada

[3] Nuclear Waste Management Organization, Toronto, Canada.

[4]Commonwealth Scientific and Industrial Research Organization, Data61, Melbourne, Australia.

[5] Institute of Energy and Climate Research – Nuclear Waste Management (IEK-6) and JARA-CSD, Forschungszentrum Jülich GmbH, Jülich, Germany

*Corresponding author: Laurent.Beland@queensu.ca

**Abstract:**

This study investigates interlayer diffusion dynamics in sodium montmorillonite (Na-MMT), a smectite clay widely used in environmental remediation, pharmaceutical formulations, and advanced materials. Understanding diffusion in Na-MMT is critical, yet current models often rely on fitted parameters rather than directly linking transport to microscopic structure, and even when the structure is known, interlayer diffusion remains challenging to model. This motivates the development of a predictive, coarse-grained, geometry-based computational framework.

Our multiscale framework couples atomistic simulations with a coarse-grained mesoscale model to quantify contributions from interlayer one-, two-, and three-water pores, as well as free pores (> three-water diameter), across dry densities of 0.8–1.3 g/cm³. Experimentally derived platelet size distributions, polydispersity, and anisotropic transport behavior are explicitly incorporated.

Results indicate that interlayer pores contribute minimally to overall water diffusion at the studied densities, with transport dominated by free pores. Predicted diffusion scaling factors closely match tritium tracer measurements when interlayer throttling is included, and the model captures the pronounced anisotropy of Na-MMT. Validation against lattice Boltzmann simulations and experiments demonstrates reliable reproduction of geometric tortuosity and pore-size distributions. Despite limitations, including rigid platelets and omission of three-water energy minima, the coarse-grained framework provides a robust platform for understanding nanoconfined diffusion, with future work aimed at refining interlayer energy landscapes and incorporating flexible platelet mechanics.

## Introduction

The transport of water and ions in charged hydrated nanomaterials is a complex, multiscale phenomenon with implications for environmental science, pharmaceuticals, and advanced materials. (Luckham and Rossi, 1999; Patra et al., 2018; Aghababai Beni and Jabbari, 2022; Liu et al., 2024) It governs key processes such as contaminant migration in groundwater, drug delivery in biological systems, and the stability of industrial formulations, including paints, coatings, and specialty composites. Beyond its practical significance in environmental and industrial applications, transport in nanoporous networks presents a fundamental challenge: do classical fluid mechanics principles hold at the nanoscale, or do new transport mechanisms emerge (Bocquet and Charlaix, 2010; Liu and Li, 2011; Kavokine et al., 2021)?

Sodium montmorillonite (Na-MMT) is a commonly employed model system for studying nanoconfined transport. It possesses a layered structure, high surface charge, and swells in the presence of water (Norrish, 1954; Chang et al., 1998; Peng et al., 2019). Na-MMT exhibits a hydration-dependent transition as its dry density increases: from a dispersed colloid to a colloidal gel, and ultimately to a nanoporous, arrested network stabilized by hydration and electrostatic forces (Paineau et al., 2011). Understanding transport across this transition is key to both fundamental and applied problems.(Choi and Oscarson, 1996; Itälä et al., 2014; Tinnacher et al., 2016) Na-MMT is a good candidate for investigating how geometric constraints and electrostatic interactions influence diffusion at the nanoscale. Furthermore, from a practical perspective, transport properties of Na-MMT are linked to its use in applications such as nuclear waste storage, where it serves as a barrier material to hinder diffusion in the clay media, or similarly in drug delivery systems, where controlled release of therapeutic agents depends on the diffusion of molecules through nanoconfined spaces.(Allen and Wood, 1988; Chrzanowski et al., 2013; Park et al., 2016)

The transport in charged hydrated nanoplatelet network is strongly influenced by particle size, shape, and surface charge. (Bradford et al., 2002; Seymour et al., 2013; Kobayashi et al., 2014; Won et al., 2019) A particularly interesting aspect of charged hydrated nanoplatelet networks is surface-induced swelling, where interparticle spacing variations create pores with distinct electrostatic fields that alter transport dynamics. Traditional colloidal interaction theories, such as the Derjaguin-Landau-Verwey-Overbeek (DLVO) model, describe long-range electrostatic and van der Waals forces but fail to capture short-range interactions, particularly in highly confined systems with interlayer spacings on the nanometer scale. (Boström et al., 2001; Raviv and Klein, 2002) In the case of Na-MMT, correctly treating hydration forces, ion correlations, and surface charge heterogeneity requires a more nuanced modelling approach.

To address these challenges, we employ a multiscale computational framework that bridges atomistic simulations with mesoscale modelling and integration of experimental data. This computational model will be used to investigate the transport properties of hydrated Na-MMT, focusing on how *interlayer pores* of one, two or three-water diameters with hindered diffusion and *free pores*, which are the pores with diameters more than three water molecules, with relatively unrestricted water-molecule diffusion affect water and ion transport at varying dry densities.

The study builds on prior all-atom molecular dynamics (MD) and coarse-grained (CG)-MD models (Zhang et al., 2022, 2023). Herein, we extend this framework by incorporating experimental platelet size distributions to account for polydispersity. A key feature of this model is the explicit inclusion of interlayer pores and free pores, which define the transport pathways. Water diffusion ($D$) in the Na-MMT is often described as a fraction, modulated by geometric constraints such as porosity ($\theta$) and tortuosity ($\tau$):

$$D = \frac{\theta D_0}{\tau} \tag{1}$$

where $D_0$ is the diffusion coefficient of free water (Oscarson et al., 1992; Briggs et al., 2017). In Na-MMT, sodium cations dissociate into a solution, generating a negatively charged clay surface that forms a dielectric double layer (DDL). (Wigger and Van Loon, 2017; Schanz and Tripathy, n.d.) While interlayer pores are known to affect ion mobility, their influence on overall diffusion remains an open question due to the challenge of capturing nanoscale heterogeneity in transport models.(Phan et al., 2020)

The CG mesoscale model comprises of 1,000 Na-MMT platelets with diameters ranging from 10 to 50 nm, spanning dry densities of Na-MMT ranging from 0.8 to 1.3 g/cm$^3$. The platelet interactions are parameterized using all-atom simulations. In contrast with conventional approaches in which porosity and tortuosity are fitted parameters calibrated against diffusion measurements, (Oscarson et al., 1992; Briggs et al., 2017) *the model of this work allows porosity and tortuosity be calculated directly.* The model will be used to verify the validity of Eq (1) at nanoscale level.

PorosityPlus and random walker codes were used to compute porosity and tortuosity, respectively, of the Na-MMT CG models (Opletal et al., 2018; Tranter et al., 2019). It was a computationally lightweight approach that was validated by comparing against computationally demanding lattice Boltzmann simulations. Our study pertains to transport in sub-100 nm charged nanoplatelet systems, a rarely addressed regime. Our findings highlight the importance or lack thereof, as we will demonstrate, of accurately accounting for nanometric interlayer contributions to transport. This study integrates atomistic precision and experimental evidence to provide a mesoscale framework for understanding nanoconfined diffusion in charged nanoplatelet systems. The results have broad implications for predictive modelling of ion and water transport in nanoclays, hydrogels, and other complex charged nanoplatelet materials.

## Methodology

### Experimental particle size determination

The Na-MMT platelet size was measured using scanning electron microscopy (SEM). Initially, Na-MMT were isolated from Wyoming MX-80 bentonite by a series of sieving, dissociation in distilled water, sonication and centrifuging steps. The isolated bentonite particles were characterized by Energy-Dispersive X-Ray Spectroscopy (EDS/EDX) to verify the nature of isolated particles. The Min and Max Feret diameters of each separated platelet or stack of platelets were measured using SEM combined with an image processing software. Electron Energy-loss spectroscopy (EELS) was used to measure the isolated Na-MMT thickness.

### Coarse-Grained Na-MMT Model

Molecular Dynamics (MD) simulations of CG Na-MMT models were performed using the LAMMPS software.(Thompson et al., 2022) The CG model is fitted to an all-atom MD simulation employing the ClayFF force field (Lammers et al., 2017; Ho et al., 2019; Underwood and Bourg, 2020). CG interaction parameters were optimized to match the hydrated potentials-of-mean force (PMF) between hydrated Na-MMT platelets generated in our previous works, which had the chemical formula of $Na_7 \cdot [Si_{84}\text{-}[Mg_7Al_{35}]\text{-}O_{192}\text{-}(OH)_{78}$ .(Zhang et al., 2022, 2023) The CG model interactions were based on a modified Morse potential for three types of interplatelet interactions: central-central, central-edge, and edge-edge. An example of coarse-graining a 2.7-nm atomistic platelet of Na-MMT to its CG counterpart is presented in Figure 1. It should be noted in the resulting CG model that water was treated implicitly, as the interaction energies were computed from *hydrated* Na-MMT systems. The optimized CG model accurately reproduces the interplatelet distances of its hydrated atomistic counterpart. Since electrostatic interactions in the atomistic system govern these distances, the CG model effectively captures their net effect. Consequently, the present CG model is specific to the Na-MMT systems from which it was derived. The technical details of this CG model optimization can be found in (Zhang et al., 2022) , and the optimized parameters of this version are listed in the Supplementary Materials.

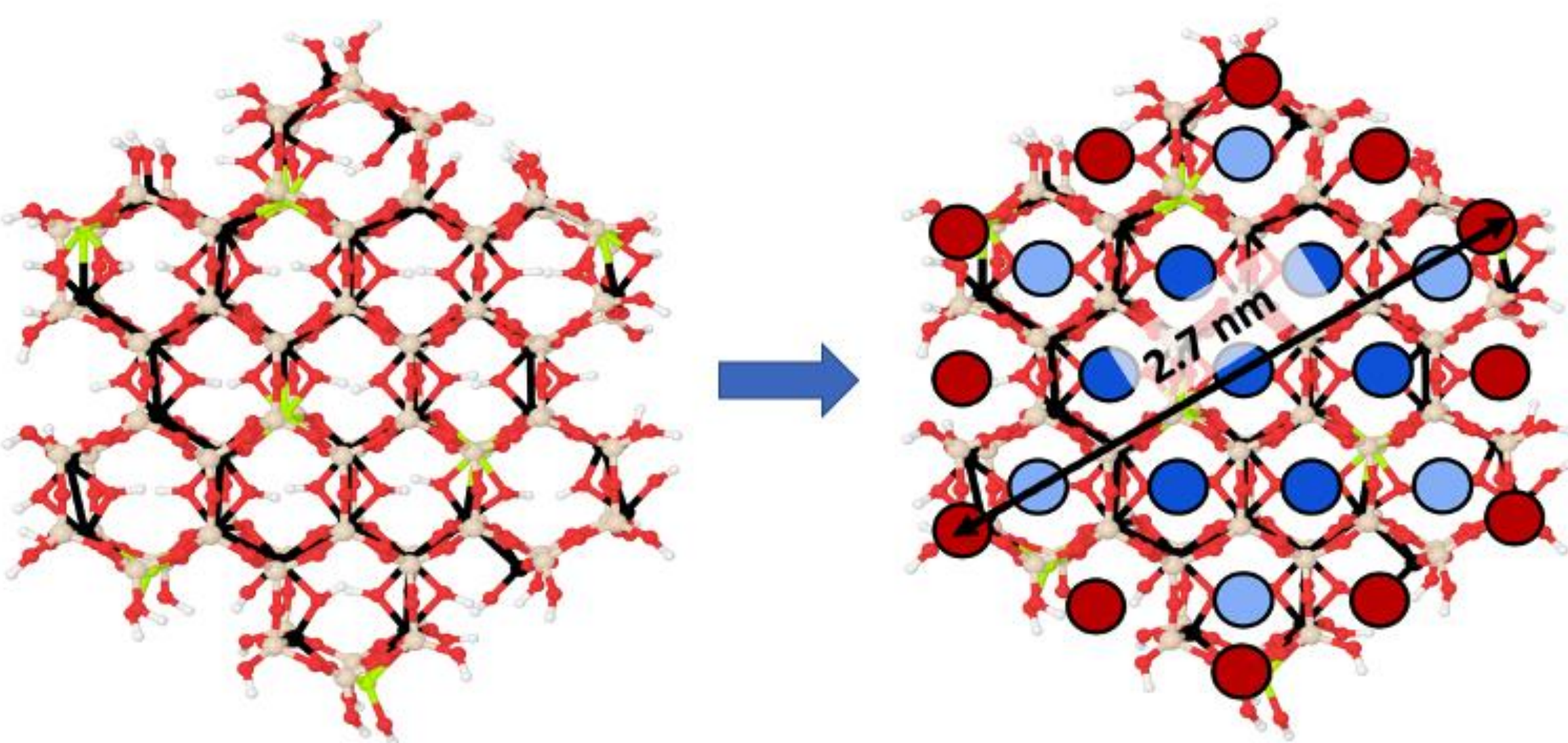

**Figure 1.** CG model based of the atomistic representation of finite Na-MMT. White, red beige, black and yellow spheres on the left represent hydrogen, oxygen, silicon, aluminium and magnesium respectively. Sodium atoms are shown but are part of the model. In the CG model, black bordered red, light blue and blue circles represent edge, central-edge and central CG particles.

Two types of simulations were created, each containing one thousand Na-MMT platelets: a polydisperse system containing 10-50 nm platelets with 5,743,760 CG particles, and an isotropic 12-nm system containing 770,000 CG particles. The particle size distribution of the polydisperse system containing 10-50 nm platelets was taken from the experimental particle size work presented in this work. The polydisperse systems were equilibrated to 300 K and 1 atmosphere, with platelets randomly placed with constraints of randomly placed with $\pm$ 20 degrees inclinations along the z-axis using Packmol to ease future compressions (Martínez et al., 2009). The polydisperse system was compressed in the $NP_zT$ ensemble with external pressure applied along the z-axis. The isotropic 12-nm systems were equilibrated to 300 K and 1 atmosphere and compressed isotopically in NPT ensemble with no constraints to preserve its isotropic nature.

Order parameters were calculated by first obtaining quaternions from the simulation and then analyzing them using the Freund package.(Ramasubramani et al., 2020) The elastic constants were calculated using the procedure used in our previous work where the systems were cooled to the 0.01 K to calculate our system's quasi-static mechanical properties by performing series of uniaxial constant pressure compression and shearing. (Ebrahimi et al., 2016; Zhang et al., 2022, 2023)

PorosityPlus was used to obtain porosity and pore size distribution for the polydisperse and isotropic systems (Opletal et al., 2018). The PorosityPlus code uses Monte Carlo integration to estimate porosity. For pore-size distribution calculations, PorosityPlus attempts to find the maximum diameter of a sphere to fill the void space. The found maximum diameter was then binned into a histogram to produce a pore-size distribution. For each network, a total of 4,000,000 insertions were performed, with each insertion attempting to move away from insertion points for 100 steps, each step having a maximum distance of 2 Å.

Once the pore size distribution (PSD) is obtained, the systems are prepared for tortuosity analysis using random walk simulations. A snapshot of Na-MMT model was converted into a discretized mesh grid for random walkers. The random walkers propagate on a discretized mesh in which each voxel is assigned a value between 0 and 1, where 0 denotes solid (non-pore) regions and 1 denotes accessible pore space. Random walkers can move only through voxels with non-zero pore values. The tortuosity between two points is then quantified as the ratio of the mean path length of the random walkers to the corresponding Euclidean (shortest) distance between them. A Pytrax code using thousands of random walkers were used to determine the tortuosity of CG models.(Tranter et al., 2019).

The hindered interlayer diffusion mechanism reported by Bourg and Sposito was incorporated into our model using a modified Pytrax random walker code (Bourg and Sposito, 2010; Tranter et al., 2019). This modification allows partial barriers to be represented using fractional voxel values ranging from 0 to 1. Each fractional voxel acts as a probabilistic filter, where the fraction value represents the likelihood of a random walker successfully transitioning to the adjacent voxel in a given direction. Consequently, a walker at a voxel with 0.8 has an 80% probability of propagating forward and a 20% probability of being reflected (i.e., remaining in its current position for that iteration).

Following the approach of Bourg and Sposito (Bourg and Sposito, 2010), we describe hindered water diffusion in interlayer pores using the normalized coefficient $D/D_0$ (where $D_0$ is the free diffusion coefficient). However, whereas Bourg and Sposito assigned a single, average $D/D_0$ value to each discrete water-layer state (e.g., 0.05 for one layer, 0.27 for two layers), we posit that diffusion should be dependent on the distance from the clay platelet surface(Rotenberg et al., 2010) . A hybrid interlayer pore model was developed. In our hybrid model, the zone immediately adjacent to the platelet surface is assigned the maximum hindrance ($D/D_0 = 0.05$), with values increasing stepwise to 0.27 and 0.44 at distances corresponding to two- and three-water-layer interlayers, respectively. This modeling approach is illustrated in Figure 2. A schematic of the conceptual hybrid model is shown on the left, contrasting it with a snapshot of its practical application in a simulated clay pore system on the right.

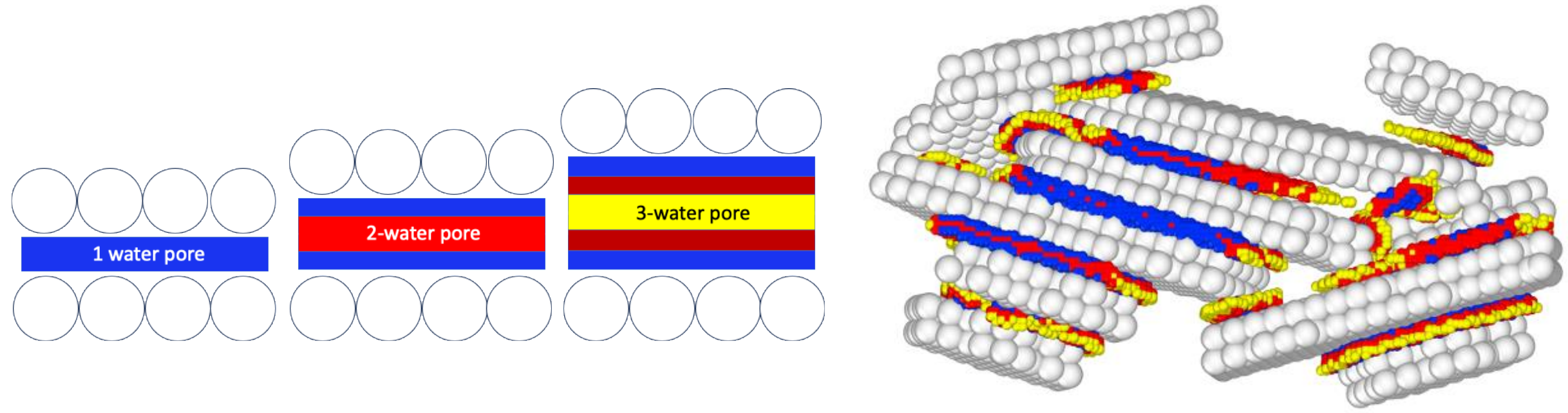

**Figure 2.** Different interlayer pores. Blue, red, and yellow correspond to one, two, and three-water interlayer pore meshes. Large white spheres correspond to Na-MMT grains.

To systematically quantify diffusive transport within the interlayer, we modelled three distinct regimes of random walkers, each representing a specific physical hypothesis:

a) *Interlayer No Throttle* model*:* Interlayer diffusivity was equated to the free pore diffusion. This model provides a theoretical baseline to isolate effect of pure geometric confinement.
b) *Interlayer Throttled* model was parametrized using interlayer random walk probabilities (0.05, 0.27, and 0.44 for one-, two-, and three-water molecule pores, respectively) derived from all-atom molecular dynamics simulations (Bourg and Sposito, 2010).This regime represents the atomistically-derived hindered interlayer diffusion.
c) *Interlayer Forbidden* model was implemented to represent a case where pores accommodating three or fewer water molecules were deemed completely inaccessible, providing an upper-boundary case to test the system's response to the complete blockage of smaller interlayer pathways.

**Results and Discussion**

The nature of isolated clay particles were quantified using EDS/EDX and the elemental profile confirmed the particles being Na-MMT. The EDS/EDX analysis revealed a Si/Al ratio of approximately 2.3:1, consistent with reported values for montmorillonite.(*Handbook of Clay Science*, 2013) In addition, the presence of 2.2 wt% sodium and 0.9 wt% magnesium confirmed that the sample corresponds to Na-montmorillonite (Na-MMT). The detailed elemental composition of EDS/EDX is presented in Supplementary Materials. Once the nature of Na-MMT platelets were confirmed their sizes were measured. The average of  Max and Min Feret were computed as a particle size distribution. The results are presented in Figure 3, indicating a wide range of platelet size ranging from 10-500 nm. The majority of platelets were in the 10-50 nm diameter range with the mode in the 20-30 nm range.

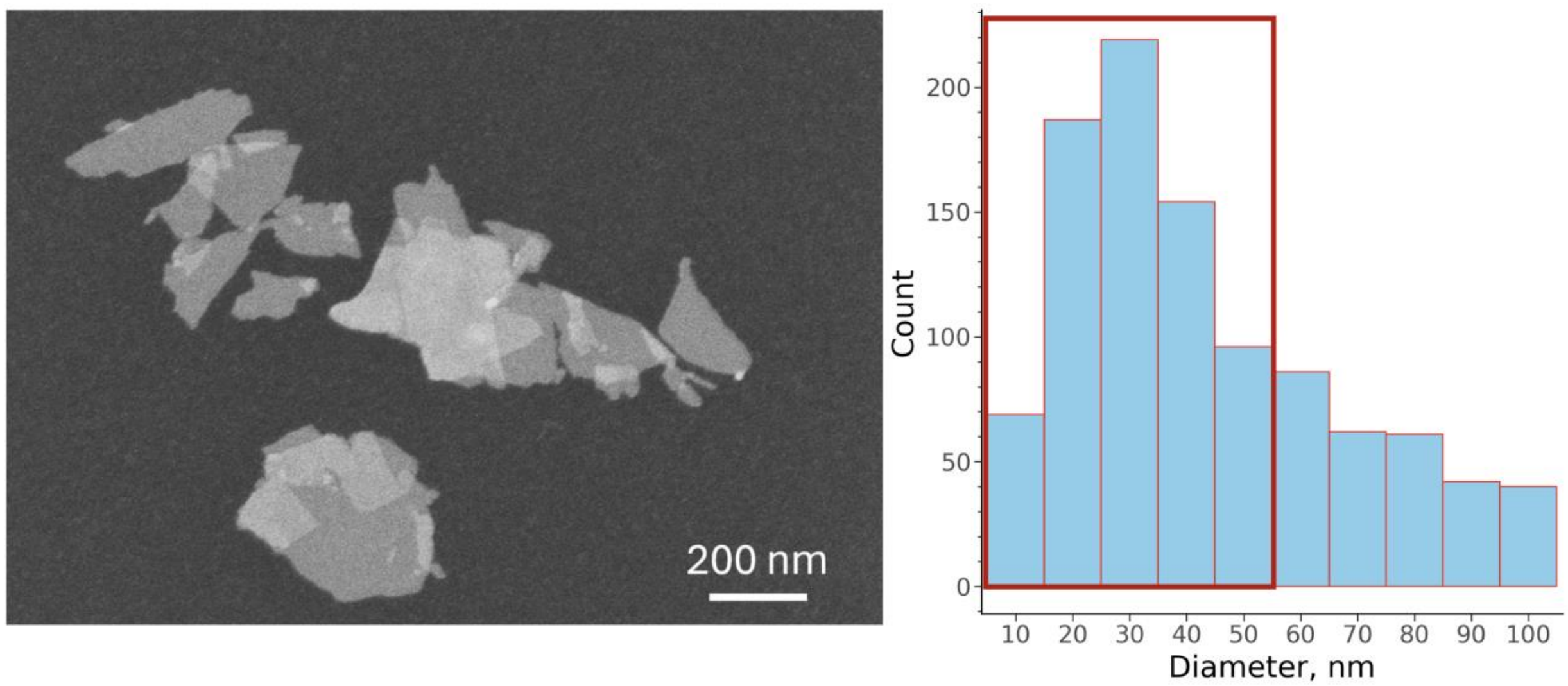

**Figure 3**. Left: Experimental scanning electron microscope image of isolated Na-MMT platelets. Right: Distribution of platelet diameters. The red square indicates the distribution used as input to the computational model.

The stacking size of the isolated Na-MMT platelets were evaluated via Electron Energy-Loss Spectroscopy (EELS) to quantify the extend of exfoliation.(Iakoubovskii et al., 2008) The measured thickness of 6-8 nm using the mean-free path of 100 nm was consistent with three to four platelets layers, assuming one nm/platelet and a hydrated interlayer with ~0.3 nm per water-molecule layer. Despite rigorous processing intended to achieve full exfoliation, the persistence of these dimensions indicates that the stacking size of three or four constituted as the thermodynamic stable unit. This finding suggested the exfoliation process has its limitations using standard dilution and centrifugation protocols.

**CG model packing and calculated elastic properties**

The CG model, using an experimental particle size distribution in the 10-50 nm range, was scaled up to a 1,000-platelet system. The initial setup was to have Na-MMT platelets randomly oriented with a 20-degree incline relative to the compaction direction (z-axis), based on the work of Underwood & Bourg, suggesting that z-axis compaction was preferred for dense, hydrated Na-MMT.(Underwood and Bourg, 2020) This orientation closely mimics the natural orientation of Na-MMT, in which gravity tends to pack the platelets along the z-axis.(Norrish, 1954; Pusch, 1992; Benson and Trast, 1995; Bobko and Ulm, 2008) The 0.8 g/cm$^3$ (dry density) system was equilibrated at 1 atm and 300K, then the simulation box was compacted in the z-direction to reach a density of 1.3 g/cm$^3$, as presented in Figure 4 resulting in six CG models with the dry densities of 0.8, 0.9, 1.0, 1.1, 1.2 and 1.3 g/cm$^3$.

The order parameter analysis showed that the system remained largely disordered along the x- and y-axes. In contrast, the order parameter along the z-axis increased from 0.89 to 0.96, approaching the value of 1 that corresponds to perfect crystalline alignment. This trend is consistent with previous all-atom simulations. (Underwood and Bourg, 2020) The observed increase is likely facilitated by the packing procedure, which imposed a 20° molecular tilt along the z-axis.

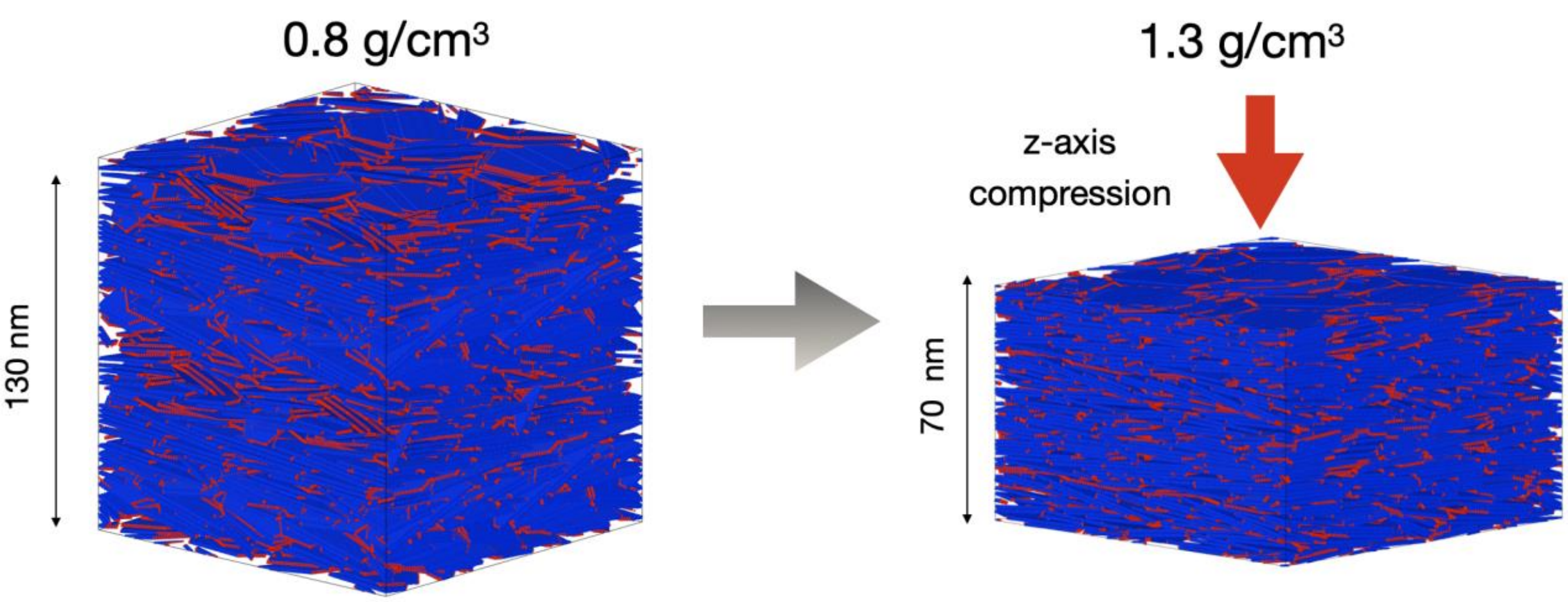

**Figure 4.** The snapshots of a polydisperse CG representation of Na-MMT platelets. A 0.8 g/cm$^3$ system (left) was compressed along the z-axis, indicated by the red arrow, to the final target dry density of 1.3 g/cm$^3$ (right). Blue and red particles represent central and edge CG particles, respectively.

The mechanical calculations revealed that the CG model was much stiffer along the plane of the platelets. The in-plane elastic constants ($C_{11}$ and $C_{22}$), which align with the platelet orientation, are up to five times larger than $C_{33}$, the out-of-plane constant. This agrees with nanoindentation measurements. (Bobko and Ulm, 2008; Sayers and den Boer, 2016) Additionally, shear elastic constants $C_{44}$ and $C_{55}$ exhibited lower shear parallel to the platelet orientation than $C_{66}$, the orientation perpendicular to it, aligning well with experimental findings. (Horseman et al., 1996; Luckham and Rossi, 1999; Sayers and den Boer, 2016)

## Calculated pore size distribution

The next step was to obtain porosity and pore size distribution (PSD) of the six 0.8-1.3 g/cm$^3$ CG systems. PorosityPlus was used to obtain these properties from still snapshots of 0.8-1.3 six g/cm$^3$ systems. The PSD was analyzed and is presented in Figure 5. Each PSD was normalized against their total volume. The most prominent feature was interlayer pores, characterized by sharp peaks below 1 nm. Free non-interlayer pores, on the other hand, exhibited a broad size distribution, with wider peaks above 1 nm, and the largest pores reaching 5 nm in the case of the 0.8 g/cm$^3$system. The broad peaks corresponded to non-interlayer pores among platelets, which were the first pores to be affected by density increase. As the system's dry density increased, the peak representing free pores gradually decreased in size. At the highest density of 1.3 g/cm$^3$, the free pore peak merged with the interlayer pore peak, indicating the consolidation of interlayer and free pore distributions.

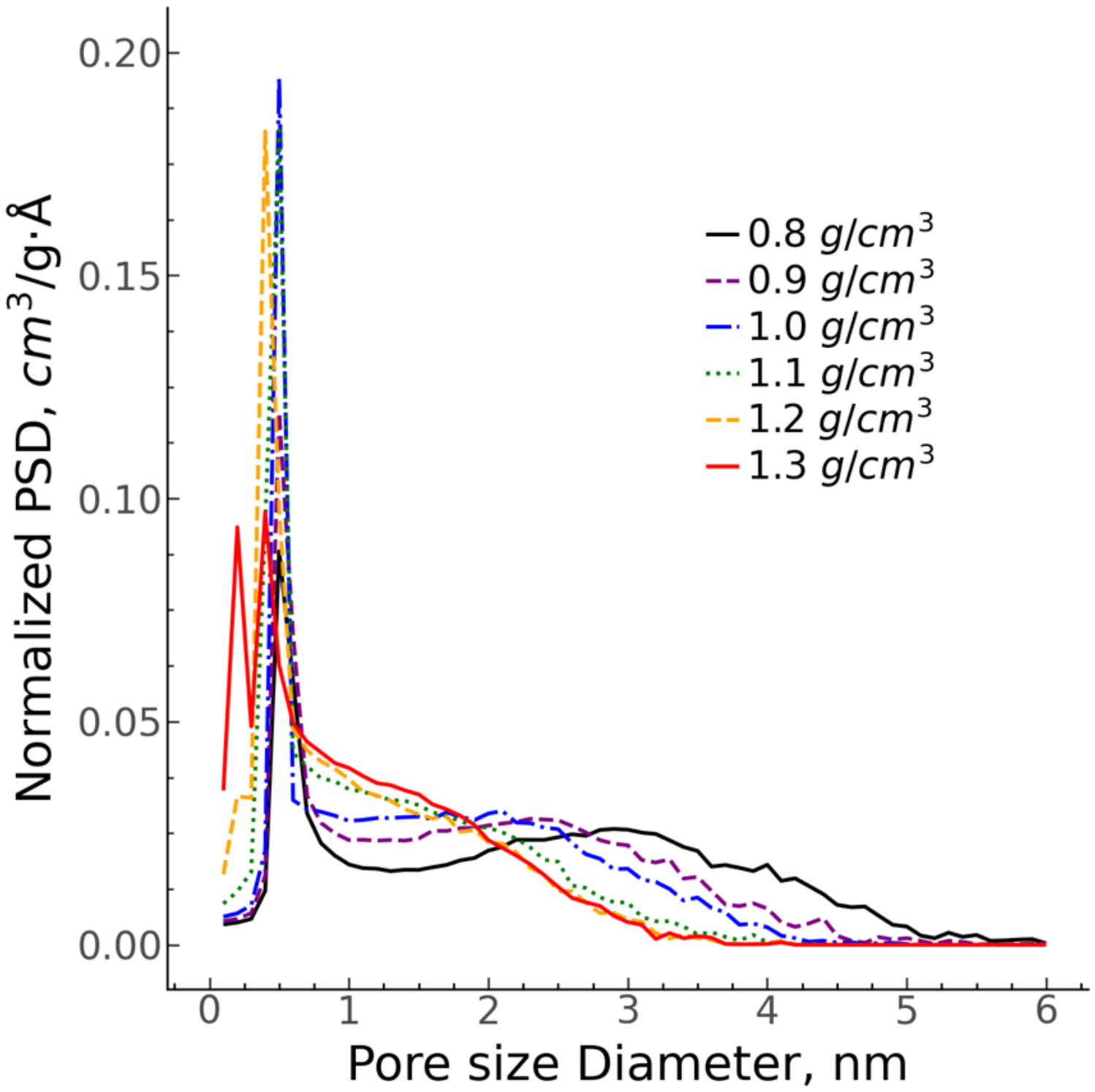

**Figure 5.** The Normalized pore size distribution of Na-MMT between 0.8-1.3 g/cm$^3$. Solid black, dashed purple, blue dot-dashed, dotted green, dashed orange and solid red lines represent 0.8, 0.9, 1.0, 1.1, 1.2 and 1.3 g/cm$^3$ dry density systems of the polydisperse Na-MMT systems, respectively.

The behaviour of interlayer pores during the compression from 0.8 g/cm$^3$ to 1.3 g/cm$^3$ revealed the gradual transition from two-water interlayer to one-water interlayer pores. At densities ranging from 0.8 to 1.1 g/cm$^3$, the dominant interlayer pores were the two-water pores, with a diameter of 0.6 nm, recalling that a single water molecule has a diameter of ~0.3 nm. At higher densities of 1.2 and 1.3 g/cm$^3$, the interlayer pore diameters shifted to smaller sizes of approximately 0.4 nm and 0.2 nm, respectively. This shift is associated with a transition from two-water pores to one-water pores. The change likely occurred because, at 1.3 g/cm$^3$, the system reached the limit of free-pore compression, forcing further compression within the interlayer spaces.

The calculated predominance of two-water interlayers aligns with experiments performed in confined systems where the relative humidity (RH) is ~80%,(Ferrage et al., 2005; Meleshyn and Bunnenberg, 2005; Gates et al., 2012) making it particularly relevant for nuclear storage applications with a target RH of 70-90%.(Villar and Lloret, 2008; Sellin and Leupin, 2013; Feng et al., 2023; "Saturation of compacted bentonite under repository conditions," n.d.) Note the absence of three-water (~0.9 nm) interlayers, which is routinely observed experimentally (Holmboe et al., 2012); this is explained by our underlying all-atom simulations not exhibiting a free-energy minimum at the three-water interlayer distance.

As the Na-MMT system underwent compression from 0.8 g/cm$^3$ to 1.3 g/cm$^3$, a predictable decrease in porosity was observed and presented in Table 1.The calculated PSD from Figure 5 was used to quantify the *interlayer* and *free pore* porosity which corresponded to pores less or equal to three water diameters ($\leq$ 0.9 nm)  and larger than three water molecule diameter > 0.9 nm, respectively.  Notably, the proportion of interlayer pores within the total pore volume increased from 8.29% to 25.45% as the system reached its target density of 1.3 g/cm$^3$. A significant trend emerged between 0.8 and 1.1 g/cm$^3$, where interlayer porosity steadily increased with compression, leading to the formation of additional interlayer pores. This trend culminated at 1.1 g/cm³, beyond which interlayer porosity remained relatively stable at approximately 0.11 and subsequently declined at 1.3 g/cm³.

**Table 1.** Total, free pore and interlayer porosity of CG Na-MMT model at various dry densities ranging 0.8-1.3 g/cm$^3$.

| **Density, g/cm$^3$** | **Total Porosity** | **Free Pore Porosity** | **Interlayer Porosity** | **Interlayer to Total Pore Ratio (%)** |
|---|---|---|---|---|
| **0.8** | 0.615 | 0.564 | 0.051 | 8.29 % |
| **0.9** | 0.565 | 0.489 | 0.076 | 13.45 % |
| **1.0** | 0.521 | 0.424 | 0.097 | 18.63 % |
| **1.1** | 0.474 | 0.363 | 0.111 | 23.42 % |
| **1.2** | 0.430 | 0.323 | 0.107 | 24.88 % |
| **1.3** | 0.385 | 0.287 | 0.098 | 25.45 % |

The observed decrease in interlayer porosity at 1.3 g/cm³ can be attributed to changes in the structural characteristics of the interlayer pores. As demonstrated in Figure 5 in the region below 10 Å, the pore size distribution at this density, represented by a solid red line, exhibits two distinct peaks rather than a single peak, indicating a transformation in the interlayer configuration. Specifically, the compression appears to induce a transition from a two-water-layer interlayer pore structure to a one-water-layer configuration. This structural modification likely accounts for the reduced interlayer porosity observed at the highest density investigated. Our CG model predicts coexistence of one-water (~0.3 nm) and two-water (~0.6 nm) interlayers, in good agreement with experimental measurements and previous calculations.(Ferrage et al., 2010)

**Random walk tortuosity Results**

The next step was to obtain tortuosity of the six 0.8-1.3 six g/cm$^3$ CG systems. The same still snapshots used for porosity calculations were used for tortuosity determination. The tortuosity measurements of the polydisperse model using random walkers were carried out in three different regimes: *No Throttle model*, representing the baseline simple model, *Interlayer Throttle model*, a model with atomistic determined interlayer diffusion resistance and *Interlayer Forbidden model*, a model testing the diffusion in case of the complete blockage of interlayer network. The results of all three regimes are presented in Table 2. The analysis considers tortuosity in two principal directions: *normal-to-compaction*, which accounts for movement in the x and y directions, and *parallel-to-compaction*, where random walkers travel along the z direction. The calculated tortuosity revealed a strong anisotropy. Parallel tortuosity were 5 to 10 times higher than their normal counterpart across all densities and densities. This indicates that diffusion was severely impeded in the direction of *parallel-to-compaction* due to all platelets were mostly aligned horizontally as presented in Figure 4.

**Table 2.** Calculated tortuosity associated with the *Interlayer not Throttled*, *Interlayer Throttled*, and *Interlayer Forbidden* scenarios in normal-to-compaction (x- and y-axis) and parallel-to-compaction (z-axis)

| Density, g/cm$^3$ | Interlayer No Throttle Normal Tortuosity | Interlayer No Throttle Parallel Tortuosity | Interlayer Throttled Normal Tortuosity | Interlayer Throttled Parallel Tortuosity | Interlayer Forbidden Normal Tortuosity | Interlayer Forbidden Parallel Tortuosity |
|---|---|---|---|---|---|---|
| **0.8** | 1.19 | 6.91 | 1.36 | 7.71 | 1.34 | 8.04 |
| **0.9** | 1.21 | 8.42 | 1.53 | 10.3 | 1.45 | 9.86 |
| **1.0** | 1.22 | 9.82 | 1.73 | 12.9 | 1.57 | 11.7 |
| **1.1** | 1.29 | 12.2 | 1.97 | 17.0 | 1.73 | 14.2 |
| **1.2** | 1.37 | 13.9 | 2.09 | 18.3 | 1.82 | 15.2 |
| **1.3** | 1.48 | 15.3 | 2.29 | 19.8 | 1.98 | 16.0 |

The tortuosity analysis revealed that the impact of nanoconfined physics was amplified by increasing the system density. A low density all models had relatively similar tortuosity. As the system density increased, the models' behaviour started to diverge. As expected, the *Interlayer No Throttle* model was the fastest model and had the lowest tortuosity values, because every pore had the maximum diffusion rate. However, the *Interlayer Throttled* and *Interlayer Forbidden* models had the crossover results. Intuition suggest that the complete interlayer pore blockage would further hinder the diffusion, our results showed the opposite at higher densities: the *Interlayer Forbidden* model yielded lower tortuosity than the *Interlayer Throttled* model. This implies that at certain density, it was more efficient for diffusing species to be completely excluded from the interlayer pore than being permitted to access and be temporarily trapped in them. The *Interlayer Throttled* model forced random walkers to spend more time in high-resistance interlayer pores, whereas *Interlayer Forbidden* model reinforced a more direct, despite being more constrained, path through the larger non-interlayer pore network.

**Diffusion Scaling factor**

With the porosity and tortuosity values determined from the previous two section, the diffusion scaling factor (*$D/D_0$)* was calculated using Eq (1). The results from random walk simulations for *No Interlayer Throttle* model and compared against independent benchmarks: pore scale Lattice Boltzmann (LB) simulation (Yang and Wang, 2018, 2019; Yang et al., 2024) and experimental tritium water diffusion data in montmorillonite or -Kunipia-F bentonite, known to the extremely high montmorillonite (~99%) content is presented in Figure 6. (Kozaki et al., 1999; Nakazawa et al., 1999; Sato and Suzuki, 2003; Suzuki et al., 2004)

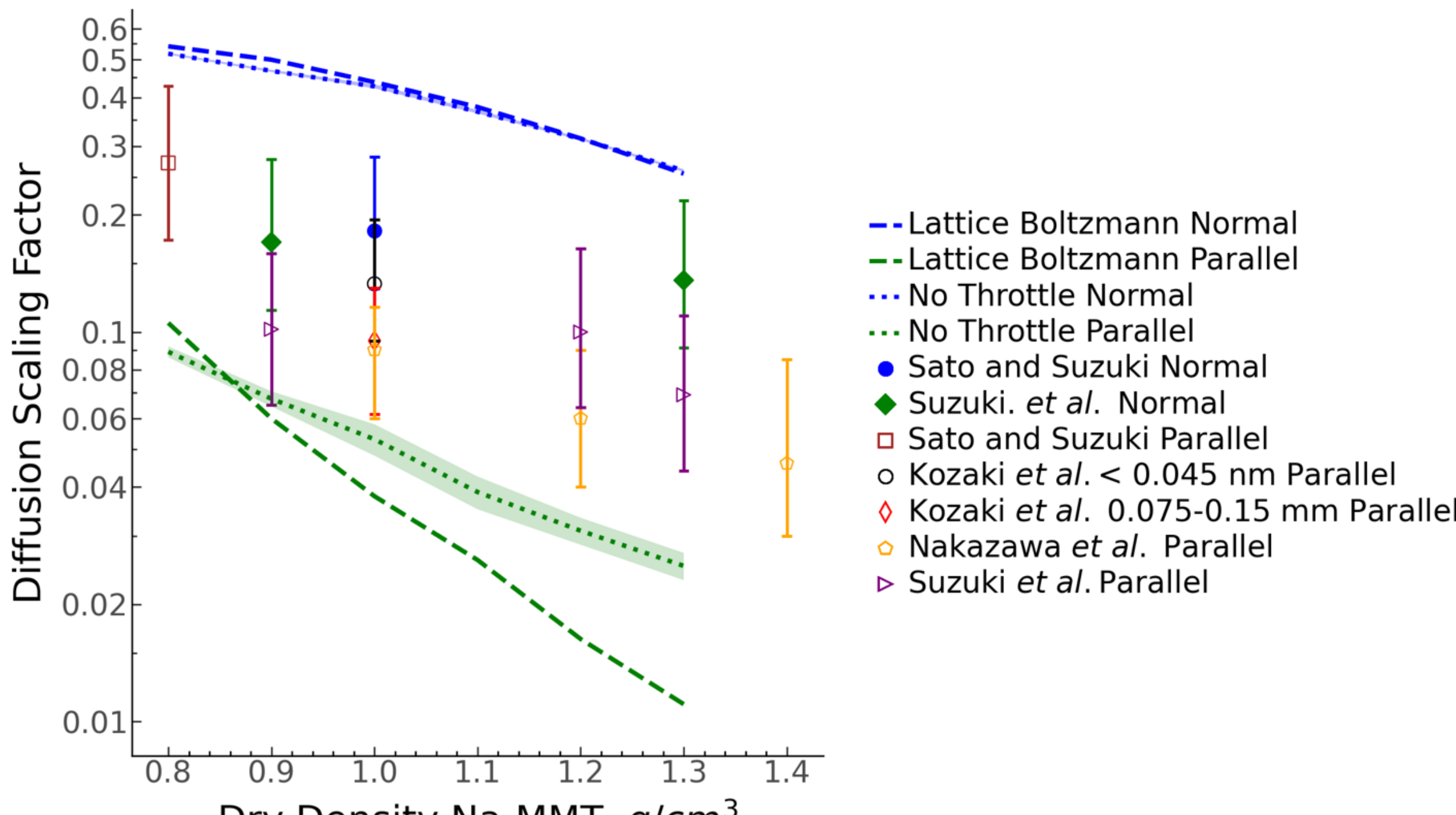

**Figure 6.** Diffusion scaling factor, $D/D_0$, of *No Interlayer Throttle*, Lattice Boltzmann (Yang and Wang, 2018, 2019; Yang et al., 2024), and experimental tritium tracer experiments of Na-MMT or Kunipia-F bentonite.(Kozaki et al., 1999; Nakazawa et al., 1999; Sato and Suzuki, 2003; Suzuki et al., 2004) The shaded blue and green regions in *No Interlayer Throttle* results indicate the standard error of the Normal and Parallel diffusion scaling factor, respectively.

The LB and *No Interlayer Throttle* model had nearly identical diffusion scaling factors. This close agreement arises because the LB method also treats interlayer pores as free diffusion pores (Yang and Wang, 2018, 2019; Yang et al., 2024). Compared with experimental data, both computational models consistently overestimate diffusion in the normal-to-compaction direction and underestimate it in the parallel-to-compaction direction. The overestimation in the normal direction is a direct consequence of both models severely under-predicting the diffusive resistance within the interlayer spaces. The systematic underestimation in the parallel direction, however, points to a more fundamental limitation of the model geometry, which will be discussed in a subsequent section.

The impact of incorporating interlayer pore models is presented in Figure 7, which compares the diffusion scaling factors for Interlayer Throttled and interlayer Forbidden systems against the experimental tritium experiments. Introducing interlayer diffusion in the *Interlayer Throttled model* significantly improved the accuracy of prediction, bridging the calculated diffusion scaling factor into closer alignment with experimental value compared to the baseline *No Throttle* model. This demonstrates that explicitly modelling the interlayer pore transport resistance is essential for accurate diffusion estimation.

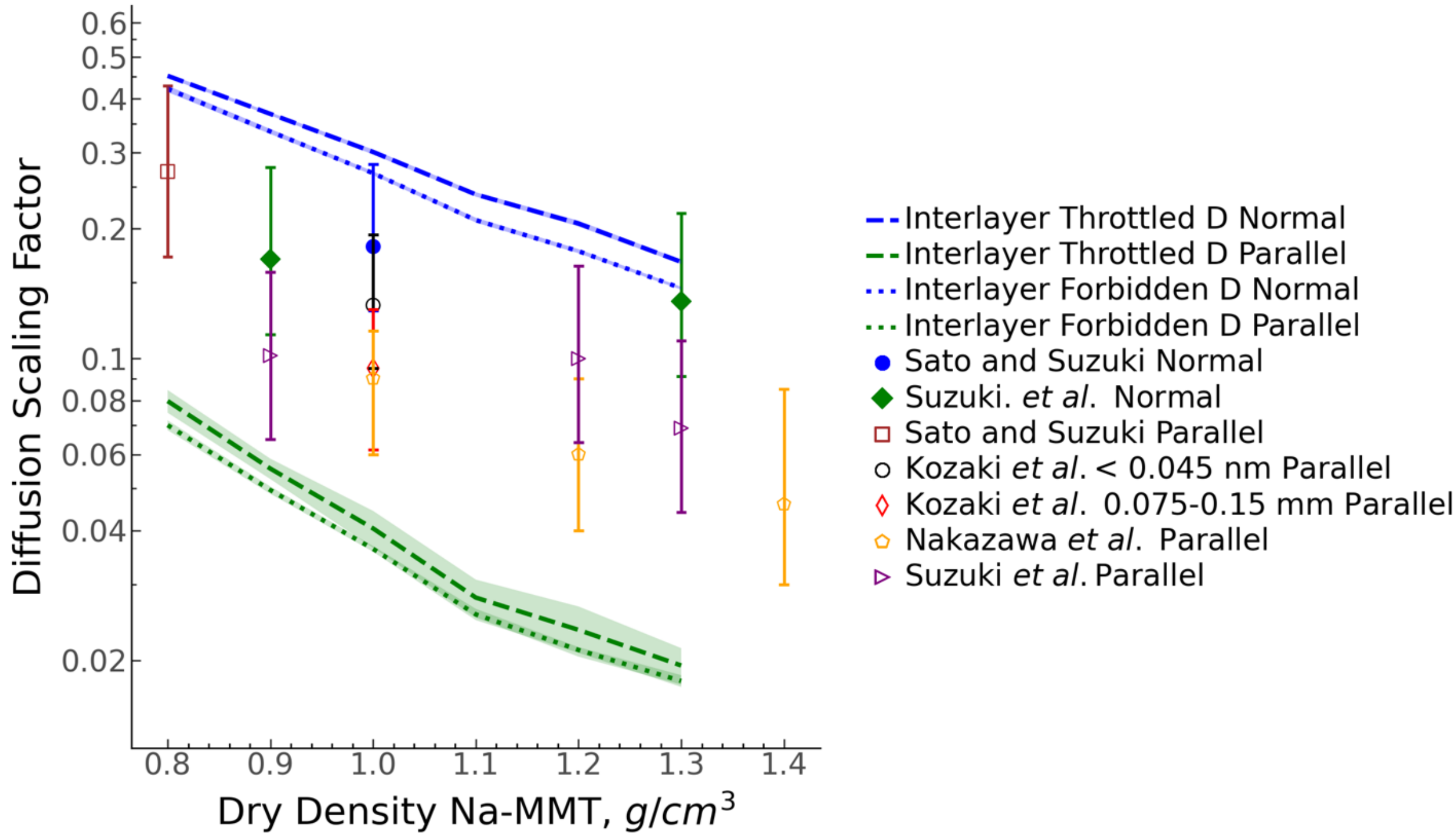

**Figure 7.** Diffusion scaling factor, $D/D_0$ of *Interlayer Throttled* (blue and green dashed lines), *Interlayer Forbidden* (blue and green dashed lines) and experimental hydrogen isotope tracer experiments of Kunipia-F bentonite presented by various single points. (Kozaki et al., 1999; Nakazawa et al., 1999; Sato and Suzuki, 2003; Suzuki et al., 2004) The shaded blue and green regions in *Interlayer Throttled* and *Interlayer Forbidden* results indicate the standard error of the Normal and Parallel diffusion scaling factor, respectively.

To further deconvolve the effects of transport resistance and porosity, we analyzed the *Interlayer Forbidden model*, where walkers are completely excluded from interlayer pores. This exclusion resulted in a diffusion scaling factor that was, at most, 20% lower than that of *the Interlayer Throttled* model. This reduction is primarily attributed to the 8–26% decrease in effective porosity from the removal of the interlayer pore network (Table 1). A key insight from this comparison is the competing effects of porosity and tortuosity: while the tortuosity of the *Interlayer Forbidden* model was actually *lower* than that of the *Interlayer Throttled* model: a factor that would typically enhance diffusion. However, this benefit was outweighed by the detrimental impact of the significantly reduced porosity. Overall, the *Interlayer Forbidden model* demonstrates that, within this system, accessible porosity is the dominant parameter governing effective diffusion, outweighing the beneficial effects of reduced tortuosity.

Our computational modeling demonstrates that when the interlayer pore volume constitutes a minority of the total porosity (less than 26%), its contribution to the overall water diffusion in hydrated Na-MMT becomes secondary. In this regime, diffusion is predominantly governed by the network of free pores, as evidenced by the nearly identical scaling factors predicted by the *Interlayer Throttled* and *Interlayer Forbidden* models. This indicates that while the physical process of diffusion within individual interlayers is highly throttled, the net impact of this entire pore network on the macroscopic transport property is negligible for studied CG Na-MMT models. Therefore, for the systems studied, the dominant factor controlling water diffusion is the connectivity and geometry of the free pore network, with the interlayer pores providing a negligible contribution to the total diffusive flux.

It should be noted that the diffusion rates of other species in the interlayer may vary, as Bourg and Sposito highlighted (Bourg and Sposito, 2010). Charged particles in transit will exhibit distinct Coulombic interactions within the charged interlayer environment, which can lead to diffusion delay or acceleration, but our study did not account for this effect. (Yang and Wang, 2018, 2019) Furthermore, the inclusion of different counterions (e.g., $Ca^{2+}$ vs. $Na^{+}$) is known to alter the energy profile of interplatelet interactions, which in turn modifies the global diffusion pore network (Shen and Bourg, 2022; Pedram et al., 2025) This microstructural dependence is clearly demonstrated in the experimental work of Choi and Oscarson, who showed that $Ca^{2+}$-saturated bentonite exhibits total intrinsic diffusion coefficients 2 to 6 times higher than its $Na^{+}$-saturated counterpart for $I^{-}$, $Sr^{2+}$, and tritiated water. (Choi and Oscarson, 1996) To model such scenarios, the present coarse-grained framework would need to be rebuilt with ion-specific interaction potentials. Therefore, while our current model focuses on water diffusion using neutral tracers, future extensions that incorporate ion-specific energy landscapes, electrostatic interactions, and cation- or anion-dependent microstructural evolution will be essential to accurately predict the transport of charged species.

**Comparison to a counterfactual isotropic system**

To further investigate the impact of system anisotropy on diffusion, we compared the polydisperse Na-MMT system with a counterfactual isotropic system composed of 12-nm platelets. Constructing a polydisperse isotropic system proved challenging, due to the limited number of 50-nm platelets, constituting approximately 40% of the unit cell. The orientation of these platelets significantly influences the system's tortuosity, hindering the achievement of isotropic diffusion behaviour. To investigate the impact of the degree of anisotropy on the diffusion scaling factor, a comparison was made with a previously studied isotropic and monodisperse system composed of 12-nm platelets (Zhang et al., 2022, 2023). This comparison allowed for an analysis of how varying degrees of compaction affect the overall diffusion scaling factor in the system.

Isotropic system

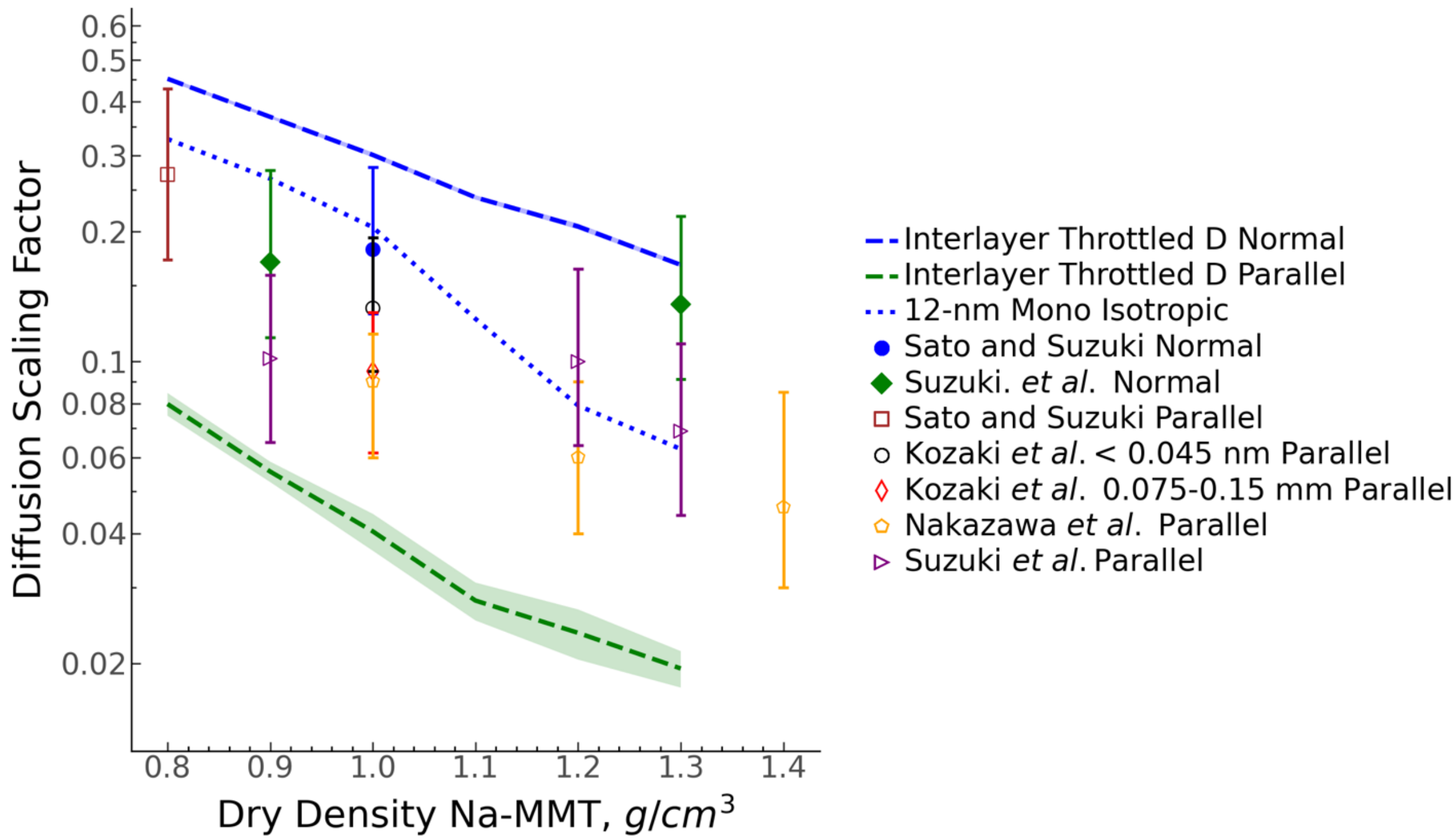

**Figure 8**. The comparison of calculated diffusion in a counterfactual isotropic system comprised of 12 nm platelets vs our polydisperse system comprised of platelets of size ranging 10-50 nm. Solid black and red triangles represent experimental tritium normal-to-compaction and parallel-to-compaction values, respectively. (Kozaki et al., 1999; Nakazawa et al., 1999; Sato and Suzuki, 2003; Suzuki et al., 2004)

The comparison of diffusion scaling factors between the isotropic monodisperse Na-MMT and z-compressed polydisperse Na-MMT is illustrated in Figure 8. Isotropic diffusion exhibited uniformity in all directions, confirming proper isotropic behaviour. The isotropic diffusion values fell between those of normal-to-compaction and parallel-to-compaction diffusion. At low densities, the isotropic $D/D_0$ values aligned closely with the experimental normal-to-compaction values, whereas at high densities, they converged toward the parallel-to-compaction experimental values. This suggests that the model's normal-to-compaction tortuosity at low densities should approximate the corresponding isotropic values. Similarly, the tortuosity of the high-density parallel-to-compaction system should align more closely with the isotropic tortuosity values

**Shortcomings of the model**

While our CG model provides valuable insights into the transport properties of Na-MMT, it is important to acknowledge its limitations and potential for future improvement. The mesoscale model demonstrates the capability to access length and time scales beyond the reach of conventional all-atom simulations. However, it remains insufficient to fully bridge the gap with continuum-scale systems. For instance, a single 100 nm platelet in this model can be represented by 56,114 coarse-grained (CG) particles. Scaling this model to simulate a system of 1,000 platelets would require approximately 56 million CG particles, demanding substantial computational resources. To enable the simulation of larger systems, additional coarse-graining strategies may be necessary to reduce computational cost while preserving the system's essential physical and structural properties.

As mentioned previously, the current Na-MMT CG model and the all-atom simulations on which it is based do not capture the three-water minima energy profile observed in several experimental works. (Holmboe et al., 2012; Muurinen et al., 2013) This could be addressed by incorporating Na-MMT energy profiles that contain such minima, such as those from Zhu *et al.* and Shen and Bourg, into the CG model. (Shen and Bourg, 2022; Zhu et al., 2022) Notably, platelet charge, size, and orientation in the all-atom calculation can have a material impact on predicted interaction free energies. (Ho et al., 2019; Pedram et al., 2025)

Another limitation of the current model arises from the rigid CG platelets. From experiments and computer simulations, it is known that Na-MMT are able to bend, which our model does not allow.(Laird et al., 1989; Appelo et al., 2010) This is likely leading to inefficient packing. Experimental studies show that larger platelets can bend and conform around smaller ones, optimizing pore space utilization. In contrast, the rigid model assumes an idealized stacking of platelets in their original, undeformed states, likely resulting in suboptimal compaction. Furthermore, the rigid model produces non-tortuous interlayer pores, which may explain the overestimation of normal-to-compaction $D/D_0$ and underestimation of parallel-to-compaction $D/D_0$. In reality, interlayer pores exhibit curvature, influencing compaction behaviour differently than predicted by the rigid model. Also, the packing process with only $\pm 20^{\circ}$ tilt of platelets may have produced a bulk system with a realistic degree of alignment which artificially favoured a single dominant orientation. More efficient compaction, including improved interlayer pore formation, would be anticipated if the platelets were flexible. Several attempts have been made to address these limitations by introducing a two-dimensional bending CG model for Na-MMT, enabling a more realistic representation of platelet deformation and compaction behaviour.(Honorio et al., 2018; Ghazanfari et al., 2022) The bending CG models were able to capture the bending modulus and interlayer shear behaviour of Na-MMT.

Our model assumes idealized hexagonal platelet geometry as presented in Figure 1, whereas SEM characterization (Figure 3) reveals that Na-MMT platelets in reality exhibit irregular, polydisperse shapes. Such morphological variation could influence local packing, porosity, and diffusion pathways, particularly under anisotropic compaction. Additionally, all underlying all-atom simulations were performed at one atmosphere water pressure and 300 K. In practice, interplatelet interaction and by extension the CG interaction potentials, are expected to evolve with changes in temperature and water pressure, particularly under repository-relevant conditions. Future work should consider these environmental dependencies to improve transferability and robustness of the model.

The transport properties of the CG model were validated against experimental results of bentonite, which introduced additional complexities. Even though the experimental data using for work contained > 95% of montmorillonite, bentonite is still a heterogeneous clay mixture containing non-montmorillonite species (e.g., quartz, feldspar, or organic matter), which can influence its macroscopic transport and swelling properties. (Allen and Wood, 1988; Pusch, 1992) Our model focuses solely on idealized Na-MMT platelets, omitting these secondary phases and their interactions. Bentonite's hydrated structure includes not only the interlayer nanopores captured by our model but also macropores - micron-scale voids between clay aggregates that significantly influence fluid transport and swelling behaviour. The absence of these macropores in our simulations may lead to discrepancies with experimental data, particularly in hydration kinetics, anisotropic diffusion, and compaction behaviour, where real systems are governed by both platelet rearrangement and macroporosity collapse.

Despite its limitations, our CG model provides a useful framework for studying Na-MMT. Future work will focus on refining interlayer energy profiles and incorporating flexible platelet dynamics to enhance the model's predictive capability.

**Conclusion**

This study presents a comprehensive multiscale computational framework to investigate the transport properties of hydrated Na-MMT, a widely used smectite clay with applications in nanocomposites, environmental barriers, and biomedical systems. By integrating atomistic simulations and experimental data, we have developed a CG-MD model that accurately captures the microstructure and transport behaviour of Na-MMT under varying dry densities. Using this model, one can directly and accurately compute the Na-MMT porosity and tortuosity to estimate diffusion, yielding results qualitatively in agreement with experimental results. While interlayer pores can comprise a large fraction of total porosity, we demonstrated that they contribute minimally to overall diffusion when the interlayer pore ratio is less than 26%. 26% is the highest interlayer pore ratio we considered, obtained at a dry density of 1.3 g/cm$^3$. Instead, water diffusion predominantly occurs through more direct free pores. Also, our analysis of geometric tortuosity via random-walk simulations highlights the anisotropic nature of diffusion in Na-MMT. Diffusion in the direction parallel-to-compaction is significantly slower than in the normal-to-compaction direction, consistent with experimental observations. Including interlayer throttling in the model improves the accuracy of diffusion predictions, aligning more closely with experimental tritium tracer measurements. While the current CG model provides a useful framework for understanding nanoconfined diffusion, it has limitations, particularly in capturing the three-water minima energy profile and intra-platelet deformations. Future work will focus on incorporating more accurate interlayer energy profiles and flexible platelet dynamics to enhance the model's predictive capabilities. By bridging the gap between atomistic simulations and mesoscale transport behaviour, this work advances our understanding of nanoconfined diffusion and provides a foundation for optimizing the design and performance of materials in environmental, industrial, and biomedical applications.

**Funding Sources**

This work was financially supported by the Nuclear Waste Management Organization and the Natural Sciences and Engineering Research Council (NSERC) of Canada.

**Acknowledgment**

The authors gratefully acknowledge the Digital Research Alliance of Canada (formerly known as Compute Canada) and the Centre of Advanced Computing at Queen's University for the generous allocation of computing resources. Special thanks to Dr. Tom G. Tranter for providing a modified Pytrax code for this study.